\documentclass[twocolumn]{aastex62}

\newcommand{\kms}{km$\;$s$^{-1}$}
\newcommand{\cms}{cm$\;$s$^{-1}$}
\newcommand{\Bf}{{magnetic field}}
\newcommand{\Bfs}{{magnetic fields}}
\newcommand{\Alfven}{Alfv\'{e}n }

\received{December 4, 2018}
\accepted{December 27, 2018}
\submitjournal{The Astrophysical Journal}

\shorttitle{Prominences in AR Sco}
\shortauthors{Garnavich et al.}

\begin{document}


\title{Driving the Beat: Time-Resolved Spectra of the White Dwarf Pulsar AR~Scorpii}


\correspondingauthor{Peter Garnavich}
\email{pgarnavi@nd.edu}

\author{Peter Garnavich}
\affiliation{Department of Physics, University of Notre Dame, Notre Dame, IN 46556, USA}

\author{Colin Littlefield}
\affiliation{Department of Physics, University of Notre Dame, Notre Dame, IN 46556, USA}

\author{Stella Kafka}
\affiliation{American Association of Variable Star Observers, Cambridge, MA}

\author{Mark Kennedy}
\affiliation{University of Manchester, Manchester, UK}

\author{Paul Callanan}
\affiliation{Department of Physics, University College Cork, Cork, Ireland}

\author{Dinshaw S. Balsara}
\affiliation{Department of Physics, University of Notre Dame, Notre Dame, IN 46556, USA}

\author{Maxim Lyutikov}
\affiliation{Department of Physics and Astronomy, Purdue University, West Lafayette, IN, 47907 USA}



\begin{abstract}

We obtained high temporal resolution spectroscopy of the unusual binary system AR~Sco covering nearly an orbit. The H$\alpha$ emission shows a complex line structure similar to that seen in some polars during quiescence. Such emission is thought to be due to long-lived prominences originating on the red dwarf. A difference between AR~Sco and these other systems is that the white dwarf in AR~Sco is rapidly spinning relative to the orbital period. "Slingshot" prominences stable at 3 to 5 stellar radii require surface magnetic fields between 100 and 500~G. This is comparable to the estimated WD magnetic field strength near the surface of the secondary. Our time-resolved spectra also show emission fluxes, line equivalent widths, and continuum color varying over the orbit and the beat/spin periods of the system. During much of the orbit, the optical spectral variations are consistent with synchrotron emission with the highest energy electrons cooling between pulses. On the time-scale of the beat/spin period we detect red and blue-shifted H$\alpha$ emission flashes that reach velocities of 700~\kms .  Red-shifted Balmer emission flashes are correlated with the bright phases of the continuum beat pulses while blue-shifted flashes appear to prefer the time of minimum in the beat light curve.  We propose that much of the energy generated in AR~Sco comes from fast magnetic reconnection events occurring near the inward face of the secondary and we show that the energy generated by magnetic reconnection can account for the observed excess luminosity from the system.

\end{abstract}





\section{Introduction}

The unique variable AR~Scorpii (AR~Sco) displays large-amplitude, highly periodic pulsations across the electromagnetic spectrum on a time scale of minutes \citep{marsh16, takata18, marcote17,stanway}. A slower modulation at the binary system's 3.56-h orbital period is also readily seen in the light curves \citep{marsh16, stiller18}. The strongest pulses of optical light are seen at a period of 1.97 minutes and appear to be a side-band or beat between the 1.95 minute spin period of the white dwarf (WD) primary and the orbit \citep{marsh16}.  Unlike typical cataclysmic variable stars (CVs), there appears to be little accretion on to the WD from its M-dwarf companion \citep{marsh16, takata17}, meaning gravitational potenitial energy is not the source of the emission. Instead, AR~Sco has been called a white-dwarf pulsar because its strong light variations consist of synchrotron radiation apparently powered by the spin-down of its highly magnetized ($\lesssim$~500~MG) WD, similar to neutron-star pulsars \citep{buckley17}. The spin-down rate has been challenged by \citet{potter18a}, placing the WD-pulsar model in jeopardy. However, \citet{stiller18} used a three-year baseline to measure a spin period decay rate that can provide more than enough energy to account for its luminosity. 

Exactly how the spin energy is converted into radiation remains uncertain. \citet{takata17} and \citet{buckley17} have postulated that the pulsed emission at the beat period is generated through a magneto-hydrodynamic (MHD) process near the secondary star. The magnetic interaction may occur with the red star's atmosphere \citep{katz17} or its wind \citep{geng16}. Electrons accelerated in the beat interaction may end up trapped in the WD magnetosphere and continue to generate synchrotron radiation for long periods of time \citep{takata17,takata18}. Detailed polarimetric observations made by \citet{potter18b} suggest that the polarized light comes directly from the magnetic poles of the WD.

CVs generally consist of a WD accreting gas from a cool dwarf. Their optical spectra tend to be dominated by light from accretion structures (disks for systems with low magnetic fields; curtains and columns for systems with high field strengths). In long orbital period systems, the spectrum of the secondary can contribute significantly at red wavelengths. The AR~Sco spectra from \citet{marsh16} shows a blue continuum added to the red spectrum of a M5-type dwarf \citep{marsh16}. In addition, strong Balmer and neutral helium emission lines are present. Their velocity variations suggest that the line emission originates from near the surface of the M~dwarf star facing the WD. The radial velocity variation of the secondary tell us when the two stars of the binary are aligned \citep{marsh16}. Zero orbital phase is at inferior conjunction, defined as when the secondary is closest to us. The minimum of the optical light curve occurs slightly later than zero orbital phase and the optical maximum occurs slightly before phase 0.5. Even after removing the effects of the pulsed radiation from the light curve, the times of minimum and maximum are offset from the orbital alignment \citep{stiller18}.

\begin{figure}
\includegraphics[width=0.5\textwidth]{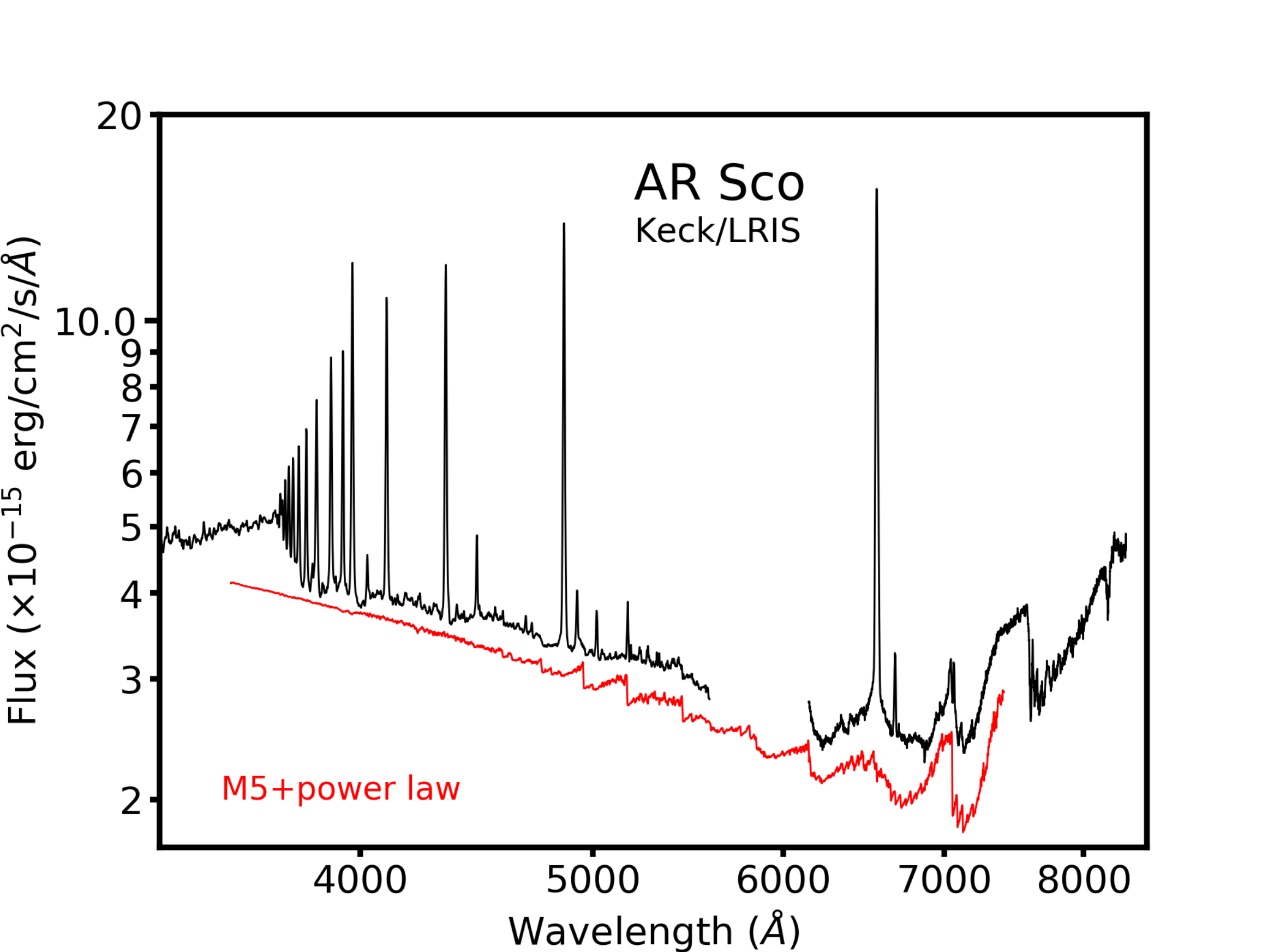}
\caption{The average spectrum of AR~Sco over three hours of observation. The gap around 5900~\AA\ is due to the dichroic dividing the light into the two arms of LRIS. The spectrum is dominated by Balmer and neutral helium emission lines. In the red, broad spectral features of the M-dwarf are evident. Ca~II H\&K emission lines are comparable in strength with the Balmer lines. The red line below the AR~Sco spectrum shows an M5 star spectrum added to a smooth power-law continuum. }
\label{average_spec}
\end{figure}


Here we present moderate spectral resolution, time-resolved spectroscopy of AR~Sco that shows significant structure in the Balmer emission lines as they vary over the orbit of the binary. We apply tomographic techniques to determine where the emission originates in the system. We discuss the impact of these structures on possible models for the source of energy driving the unusual emission in AR~Sco. 

\section{Data}

We obtained spectra of AR~Sco with the Low-Resolution Imaging Spectrograph \citep[LRIS;][]{oke95} on the Keck-I telescope on 2018 March 15 (UT). The red and blue arms of the spectrograph were divided using the 560~nm dichroic. On the blue side we employed the 600/4000 grism, while on the red arm we used the 900/4000 grating. The 0.7~arcsec wide slit provided a resolution (FWHM) of 2.1~\AA\ in the red and 2.8~\AA\ in the blue. 

A sequence of 15~sec long exposures was begun at 12:49 (UT) and finished at 15:54 (UT) during twilight. There is a gap in the spectral sequence lasting 15~minutes when a different target was observed. The CCD detectors were binned 2$\times$2 which reduced the readtime and overhead to approximately 1 minute. A total of 132 spectra were obtained in the red and 183 spectra in the blue. The difference is a result of the shorter overhead cost for the blue CCD.

The spectra were extracted, wavelength, and flux-calibrated using packages in IRAF\footnote{IRAF is distributed by the National Optical Astronomy Observatory, which is operated by the Association of Universities for Research in Astronomy (AURA) under a cooperative agreement with the National Science Foundation.}. The resulting average spectrum is shown in Figure~\ref{average_spec}. As
noted by \citet{marsh16}, an M5 dwarf companion is clearly visible by the presence of broad band heads and a rising red continuum. Narrow Balmer features dominate the line emission and neutral helium lines are also present. A weak HeII feature is detected and many metal lines are seen throughout the optical spectrum with the strongest being MgI at 5167~\AA. Two of the most significant emission features in the blue part of the spectrum are CaII at 3933~\AA\ and 3968~\AA. The 3968 line is comparable to, and blended with H$\epsilon$ at 3970~\AA. \citet{marsh16} noted significant IR calcium triplet emission in their spectra of AR~Sco.

The average spectrum shows a significant jump in brightness across the Balmer limit and a change in slope of the continuum. This suggests hydrogen opacity is strongly influencing the continuum, possibly from the irradiated secondary star.

\begin{figure}
\includegraphics[width=0.5\textwidth]{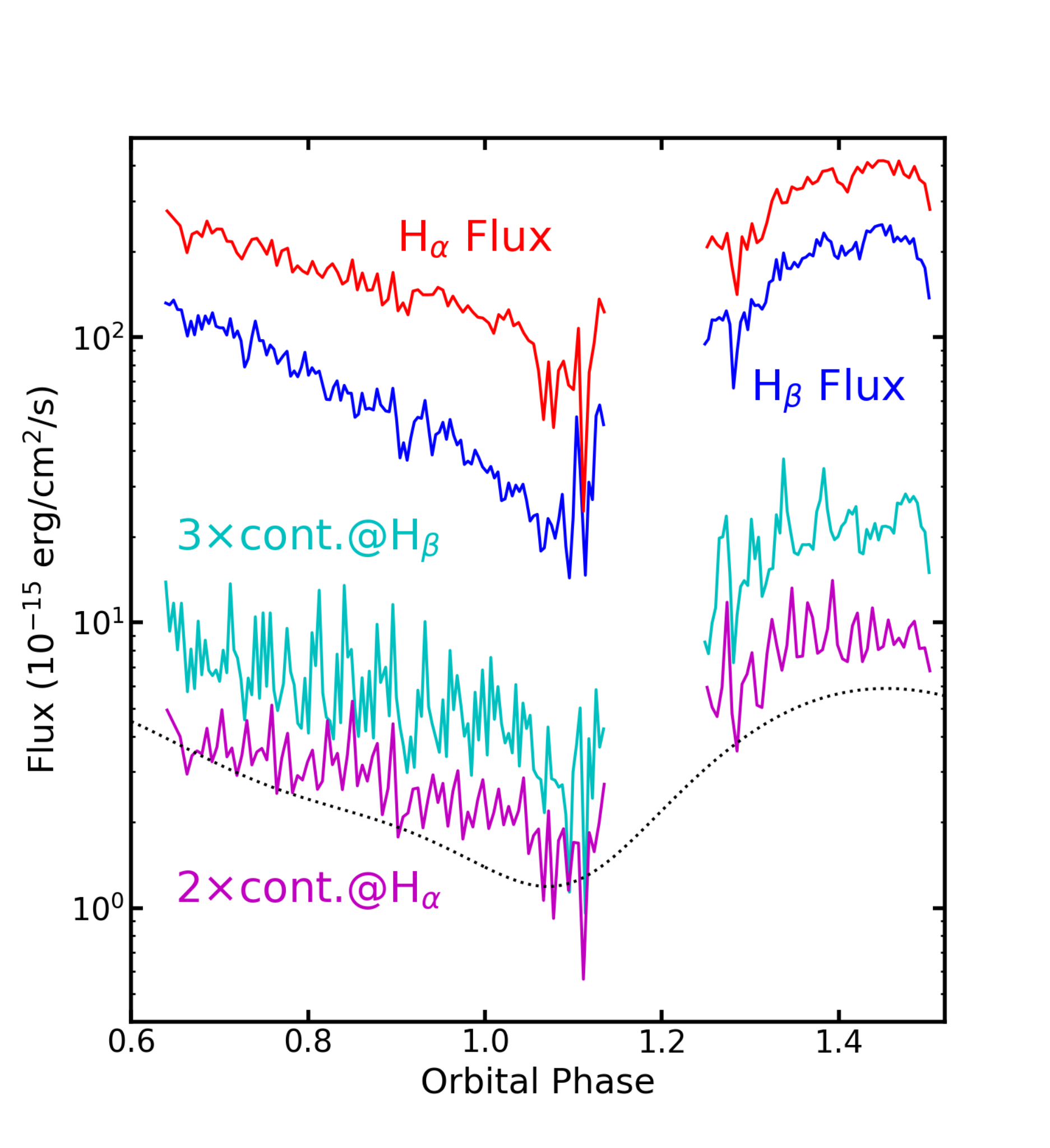}
\caption{The flux of H$\alpha$ (red line) and H$\beta$ (blue line) as a function of orbital phase. The continuum flux per \AA\ at the location of H$\alpha$ (magenta line) and H$\beta$ (cyan line) are shown with a relative shift to make the plot more compact. The beat and spin pulses are clearly seen in the continuum light curves, but are not obvious in the emission line fluxes. The dotted line is the broad-band orbital modulation of AR~Sco described by \citet{stiller18} }
\label{flux}
\end{figure}

\begin{figure}
\includegraphics[width=0.5\textwidth]{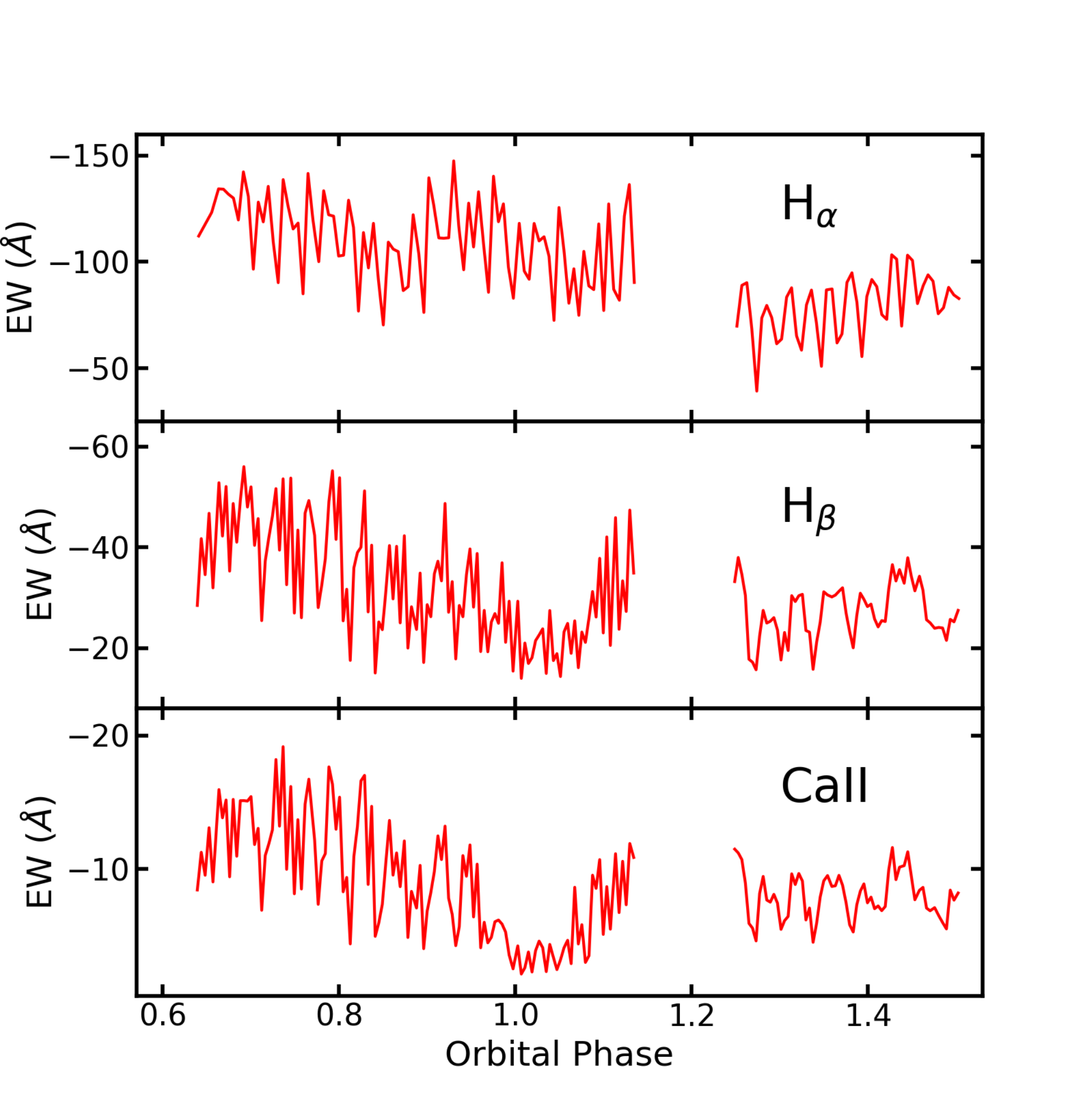}
\caption{Variability of the emission line equivalent widths (EW) over a binary orbital period. The three panels show the EW variations of H$\alpha$, H$\beta$, and, CaII 3933~\AA\ emission lines. The H$\alpha$ EW does not show a minimum around phase 1.05 because the satellite emission features significantly add to the total flux even around conjunction.
}
\label{orbit}
\end{figure}

\section{Analysis}

\subsection{Variations on the Orbital Time-scale}

Based on the orbital ephemeris by \citet{marsh16}, our Keck spectra covered nearly one orbit. There are two gaps each lasting about 10\%\ of an orbit centered at phases 0.2 and 0.6. We measured the fluxes and equivalent widths of the emission lines as well as the continua fluxes around 5300$\pm 200$~\AA\ (roughly $V$-band), 4150$\pm 150$~\AA\ (roughly $B$-band), and 3600$\pm 100$~\AA\ ($U$-band). The resulting magnitudes were then corrected to approximate the Vega system. 

\subsubsection{Emission Line Equivalent Widths and Fluxes}

Figure~\ref{flux} shows the line and continuum fluxes as a function of orbital phase. The sky conditions were clear during the observations, but seeing variations and a narrow slit width may have combined to generate false fluctuations. For example, around phase 1.1 when the star was its faintest, there are major jumps in the line and continuum fluxes that are not likely real.

The overall continuum variation and emission line flux variation is similar over the orbit. The minimum flux for the lines and continua are offset from conjunction with the secondary star and instead occur around phase 0.1. After orbital phase 0.1 the fluxes rise quickly to a peak between phases 0.4 and 0.5. There is then a slow decline back to minimum just after secondary conjunction. The light curves are similar to the orbital modulation curve derived from broad band photometry by \citet{stiller18}.

The continuum fluxes show a large-amplitude variation on a fast time-scale when compared with the line fluxes. This fast variation is a result of the beat and spin pulses and it is clear that the line emission is not strongly affected at these periods. Power spectra of the light curves show a very strong signal at the beat period in the continua, but the emission line signal is only 5\%\ of the continua amplitudes.


The emission line equivalent widths (EW) for H$\alpha$, H$\beta$, and CaII are shown in Figure~\ref{orbit}. Rapid variations due to the beat/spin pulses are seen in the EW plots and are due primarily to changes in the continuum flux. On long time-scales the H$\alpha$ EW is relatively constant between phases 0.5 and 1.1, while the H$\beta$ and CaII EW values show a clear minimum at phase 1.05 (slightly earlier than the flux measurements). A detailed look at the H$\alpha$ line shows a core with additional blue and red shifted velocity structure in the emission. These satellite structures are not as distinct in H$\beta$ and weaken quickly going to higher-order Balmer lines. The presence of these additional emission structures with a steep Balmer decrement keeps the H$\alpha$ EW relatively flat through the secondary conjunction.

\begin{figure}[h]
\includegraphics[width=0.5\textwidth]{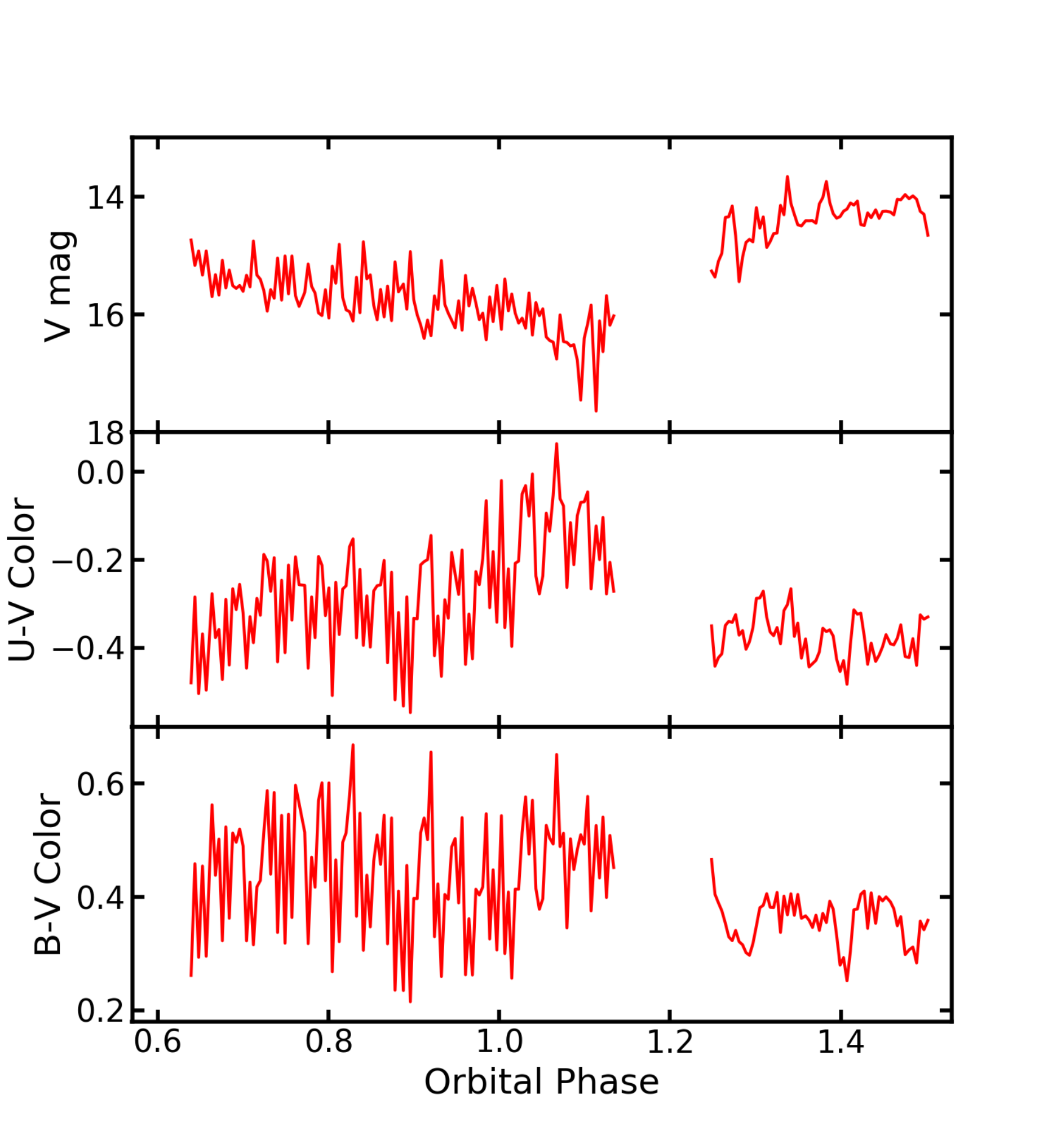}
\caption{The variation of brightness and color over an orbital cycle. {\bf top)} Approximate $V$-band magnitude synthesized from the Keck spectra. {\bf center)} The $U-V$ color synthesized from the spectra and plotted as Vega magnitudes. Large variations in color are seen at the beat/spin time scale. The $U-V$ color becomes significantly red after secondary conjunction. {\bf bottom)} The $B-V$ color derived from the Keck spectra. Here, no obvious reddening is seen after secondary conjunction.
}
\label{color}
\end{figure}

\subsubsection{Color Variations}

 The approximate $V$ magnitudes and color variations over an orbit are shown in Figure~\ref{color}. The $B-V$ and $U-B$ color curves shows significant changes at the beat/spin time-scales. Over an orbit both colors redden slightly around orbital phase 0.75 to 0.80. The $U-V$ color moves significantly to the red after conjunction, but the $B-V$ color shows, at most, only a minor change. The different behavior of these two colors may be explained if the $U$-band flux come from an area more compact than that of the $V$ and $B$-band emitting regions. Then, near conjunction, a larger fraction of the ultraviolet emitting region would be blocked by the secondary when compared with the optical bands. The system is most blue when it is brightest at orbital phases 0.25 to 0.5.

\begin{figure}
\includegraphics[width=0.5\textwidth]{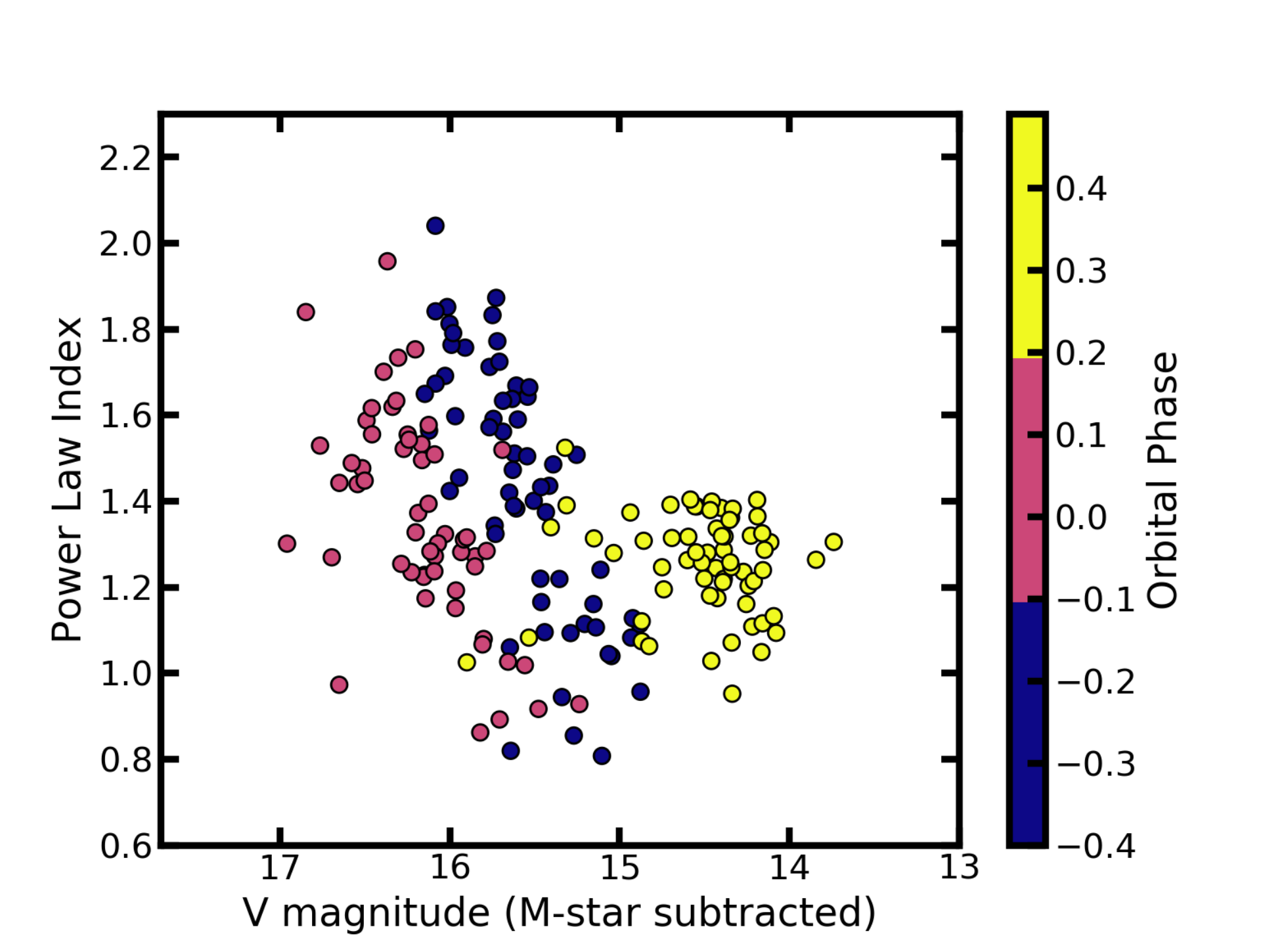}
\caption{The continuum slope versus the optical brightness of AR~Sco. After subtracting the contamination from a M5 spectrum, the slope of the optical continuum was matched to a power-law index in frequency. The color of the points indicate the orbital phase of the observation. When AR~Sco is brightest it shows little change in color. In contrast, at and prior to secondary conjunction there is an approximate linear relation between the spectral index and the magnitude. Note that synchrotron emission typically has an power-law index of 0.8.
}
\label{powerlaw}
\end{figure}

 To study how the continuum slope varies as a function of brightness, we subtracted the contamination of a M5 stellar spectrum assuming a brightness of $V=18.9$ \citep{marsh16} from each blue spectrum. Since the non-thermal spectrum is expected to primarily synchrotron emission, we converted the spectral slope after removing the secondary light into a power-law index, $s$ defined as $F_\nu \propto \nu^{-s}$. In Figure~\ref{powerlaw}, we show how the complex relation between the continuum slope and brightness evolves over an orbit period.
 
 For orbital phases between 0.2 and 0.5, the non-stellar emission is at its brightest, varying between 15$<V<$14 mag.  The optical power-law index hardly changes at this phase, remaining around $s=1.2\pm 0.2$. At orbital phases when the emission is fainter, there are significant variations in the spectral index with extreme values ranging from 0.8$<s<$2.0.
 
 For orbital phases between $-0.4$ to $-0.1$ (equivalent to phases 0.6 to 0.9), there appears a linear relationship between the power-law index and the emission luminosity. Around $V=15$ the index is a shallow $s\sim 0.9$, then steepens to 1.9 as the flux fades to $V\sim 16$. The relation between brightness and spectral slope is similar for orbital phases between $-0.1$ to 0.2 (that is, phases 0.9 to 1.2), except the emission is a half magnitude fainter while the index ranges between 0.8 and 1.8. 
 
 The spectra of many synchrotron sources in the optically thin regime are found to have power-law indices of $s \approx$0.6 to 0.8, corresponding to electron energy distributions with indices between 2.2 and 2.6 \citep{rybicki86}. The AR~Sco spectra are consistent with these spectral indices during the peaks of the beat emission. The higher energy electrons cool fastest and steepen the observed spectrum unless energy is continuously replenished. The large brightness variations observed in AR~Sco suggests that the cooling time of the relativistic electrons is shorter than the spin period of the WD \citep{takata18}. 
 
 We note that the spectral slope and system brightness are always greater that of the M5 dwarf star. It appears the synchrotron emission is always present at some level suggesting that the cooling time in parts of the system is longer than half the spin period.
 
 The spectral slopes observed as the secondary approaches superior conjunction are puzzling. The spectral index appears relatively constant, implying either the electrons have a long cooling time or their energy is continuously replenished. However, the observed slope is steeper than typical for synchrotron emission from freshly energized electrons. As the full irradiated face of the secondary is exposed around orbital phase 0.5, it may be that thermal emission from the secondary dilutes the synchrotron spectrum, or we are viewing a region of the interaction where the cooling time is long.

\subsubsection{The H$\alpha$ Trailed Spectrum}

\begin{figure}[b]
\includegraphics[width=0.5\textwidth]{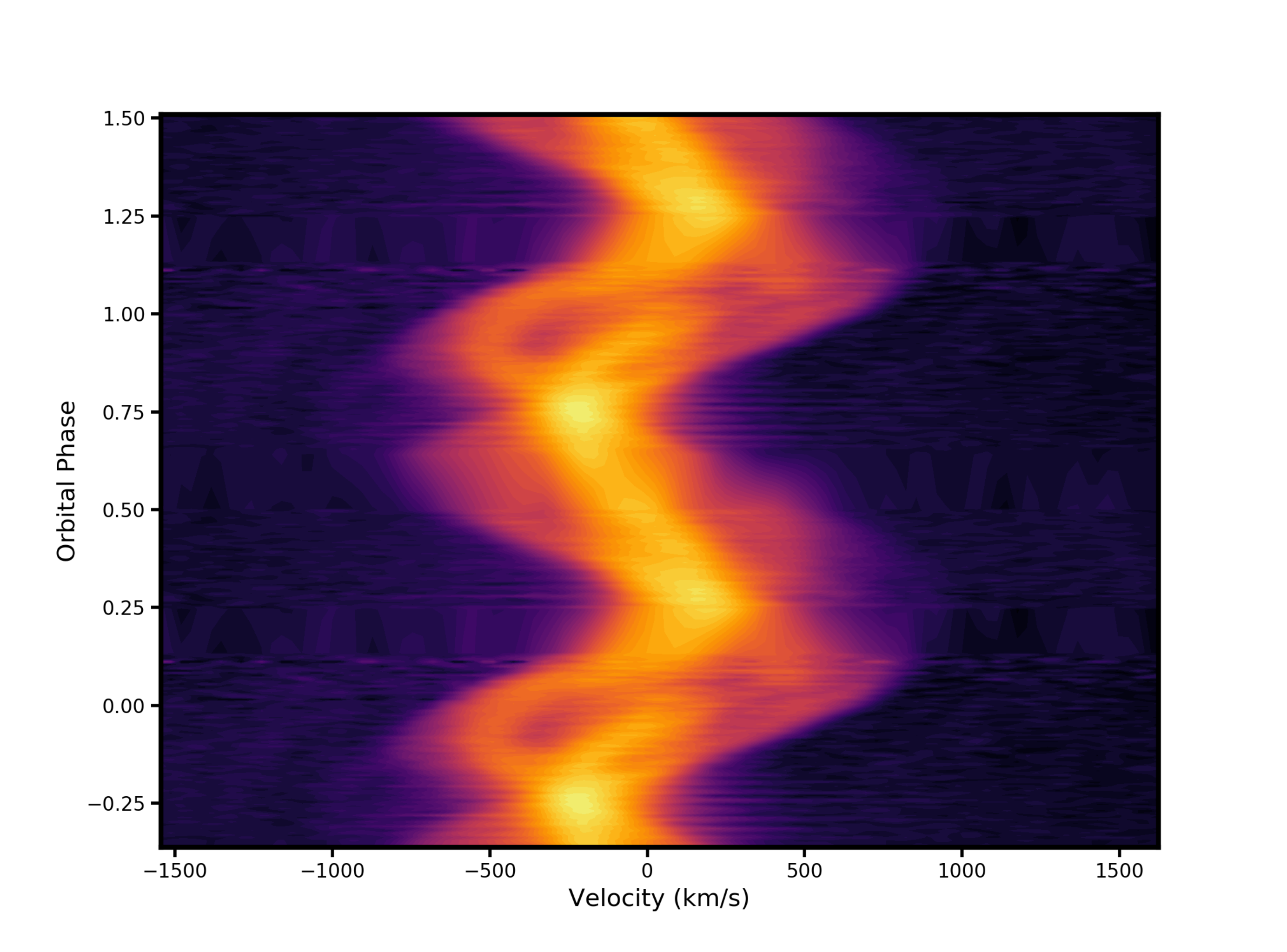}
\caption{The H$\alpha$ emission line velocity distribution as a function of orbital phase. Gaps in the phase coverage between orbital phases $\sim$0.15-0.25 and $\sim$0.5-0.7 have been linearly interpolated. The local continuum was subtracted and line was normalized so that H$\alpha$ emission has the same total flux in each spectrum. The phase coverage is repeated for clarity. }
\label{trailed_spectrum}
\end{figure}

Our spectra were obtained using a moderate dispersion grating on the red arm of LRIS and the good seeing permitted a narrow 0.7~arcsec wide slit to be employed. The resulting spectral resolution is the highest yet reported for AR~Sco. It was clear while taking the spectra that the H$\alpha$ line profiles were quite complex with multiple peaks appearing and shifting in velocity. The continuum subtracted spectra at H$\alpha$ stacked as a function of orbital phase are shown in Figure~\ref{trailed_spectrum}. Besides an ``S-wave'' from the irradiated face of the secondary star, there are satellite lines that extend out to 500~\kms\ and that are out of phase with the emission associated with the secondary. These satellite features, at times, contribute more than 30\%\ of the total H$\alpha$ flux. In fact, near phase 1.1 the emission from the irradiated face is obscured and the satellite emission lines dominate the flux.

As noted above, this satellite emission is visible in the H$\beta$ emission line, but it contributes significantly less to the total flux, suggesting a steeper Balmer decrement than for the irradiated emission.

\begin{figure}
\includegraphics[width=0.5\textwidth]{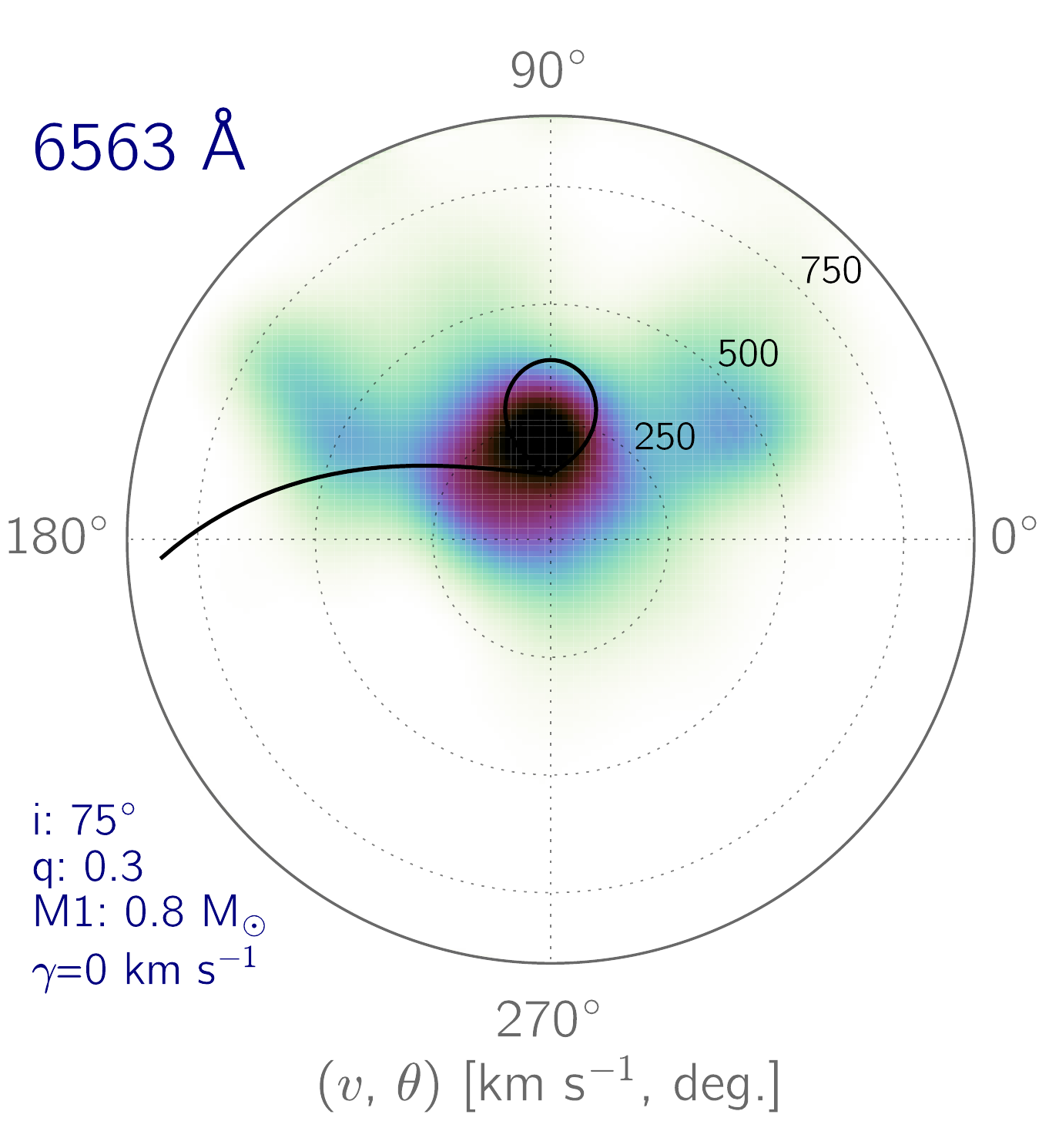}
\caption{The tomogram of the H$\alpha$ emission line showing that the central S-wave comes from the side of the red dwarf star facing the WD. The satellite emission features are centered at angular phases of 35$^\circ$ and 145$^\circ$. The velocity overlay includes both the Roche lobe of the secondary and the ballistic trajectory of the accretion stream (if it were to exist) for assumed values of the orbital inclination ($i$), the mass ratio ($q$), and the WD mass.}
\label{tomogram}
\end{figure}

A standard tomogram of the spectral series is shown in Figure~\ref{tomogram}. The majority of the H$\alpha$ emission is consistent with an origin on the face
of the secondary star while the satellite emission forms two ``Mickey Mouse ears'' at phases of  approximately 0.1 and 0.4 and extending to more than 500~\kms . This emission configuration is similar to that observed by \citet{kafka08,kafka10} in magnetic CVs during a period of low accretion rate. \citet{kafka10} attributed these satellite emission regions to a long-lived prominences (super-prominences, or slingshot prominences; \citet{collier90,slingshot}) on the secondary star. The detached post common envelope binary, QS~Vir, also shows slingshot prominences that persist for years \citep{parsons16}.

The best examples of satellite emission in magnetic systems are seen in the polars AM~Her and BL~Hyi. In polars, generally the rotation of both the white dwarf and red dwarf are synchronized to the orbit, so the magnetic interaction is fixed in the rotating frame. For AR~Sco the situation is very different with the WD and its magnetic field rotating past the red dwarf's magnetic field at the spin rate of the WD. The presence of these prominences shows a significant magnetic field on the secondary and it may be the interaction between the magnetic fields of the two stars that drives the energy production in AR~Sco.

\begin{figure}[b]
\includegraphics[width=0.5\textwidth]{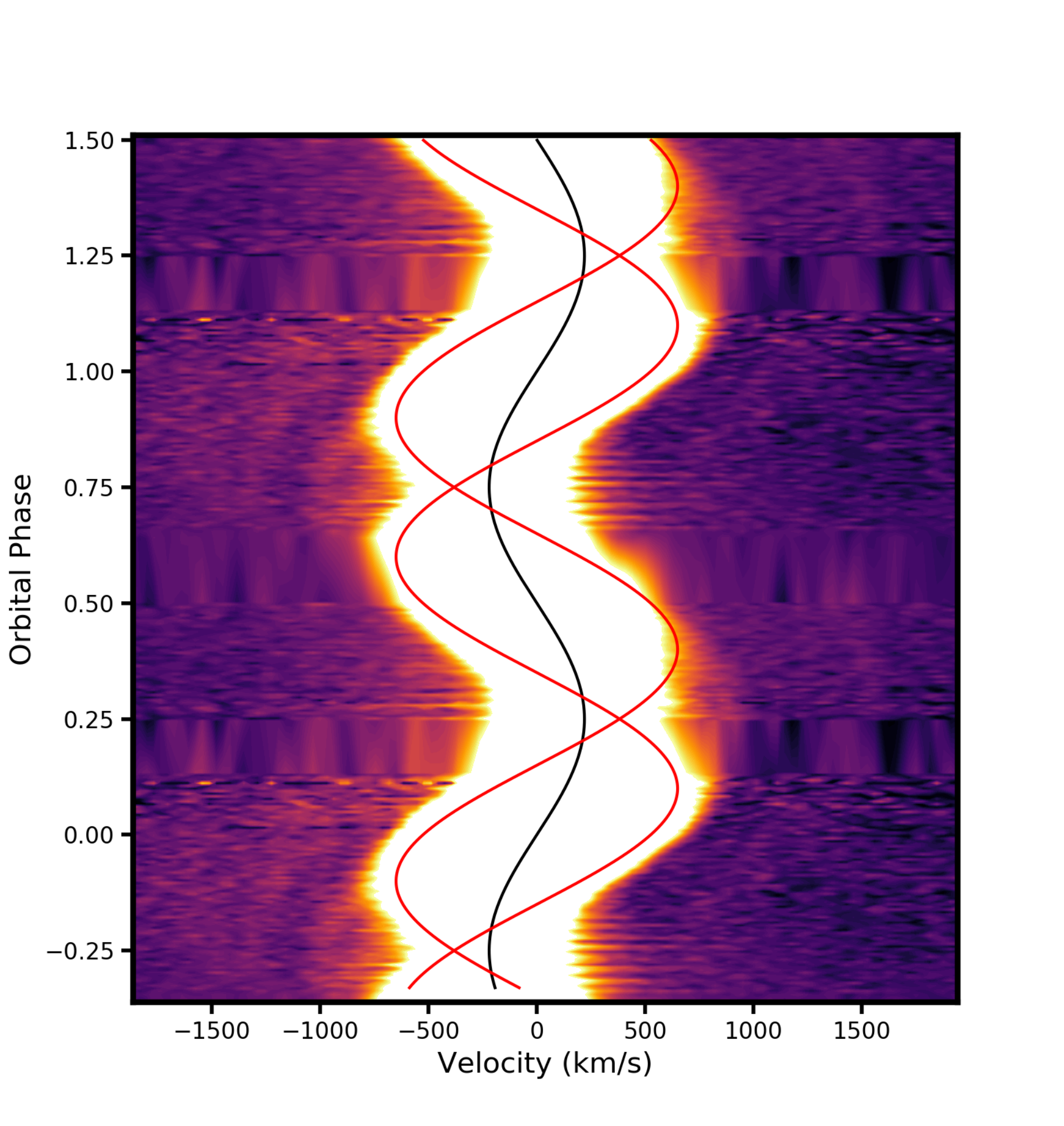}
\caption{The H$\alpha$ trailed spectrum displayed at high contrast to show the faint emission extending to 1000~\kms\ at phases 0.25 and 0.75. This emission suggests a faint prominence extends out multiple stellar radii on the side of the secondary opposite the white dwarf.  The black line show the phase and amplitude of the Balmer emission on the face of the secondary, while the red lines indicate the phase and velocity amplitude of the two bright prominences.  }
\label{prom_c}
\end{figure}

When stretching the trailed H$\alpha$ spectrum to high contrast, there appears to be weak emission out to $\sim1000$~\kms\ at phases 0.25 and 0.75 (Figure~\ref{prom_c}). This would be a faint prominence extending far from the secondary star on the side opposite the white dwarf.

\subsection{Spectral Variability at the Beat/Spin Frequencies}

\begin{figure}[b]
\includegraphics[width=0.5\textwidth]{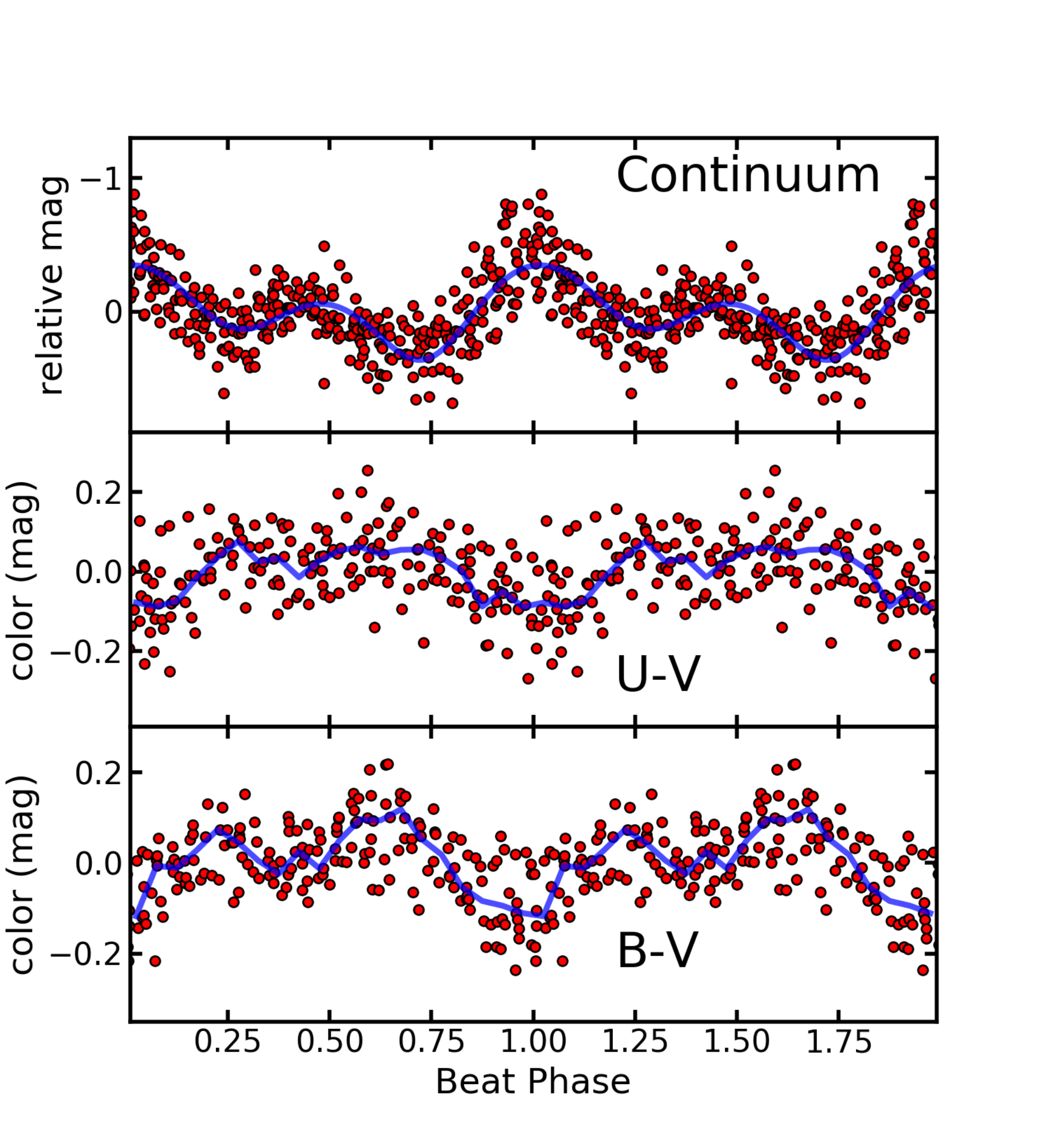}
\caption{The brightness and colors phased on the beat period. {\bf top)} The relative magnitude variations of the spectral continuum over a beat period. Red and blue continua measurements were combined to improve the time resolution. {\bf center)} The approximate $U-V$ color changes over a beat/spin period. The orbital modulations have been removed leaving only the relative color changes phased on the beat/spin period. The blue line is the data averaged into 0.05 wide phase bins.  {\bf bottom)} Same as the center panel but showing the approximate $B-V$ color variations.
}
\label{beat}
\end{figure}

To study the continuum flux changes over short time-scales, we phased the brightness and color measurements to the beat period with the result show in Figure~\ref{beat}. The double beat pulse seen in the broad-band light curves of AR~Sco \citep{stiller18} are reproduced in the spectral continua. The $U-V$ and $B-V$ colors also show a variability at the beat phase with the bluest colors occurring at the brightest peak of the main beat pulse. The reddest $B-V$ color is not seen at phase 0.75, the faintest point of the beat light curve, but shifted earlier by 0.1 in phase. Beyond the blue peak at conjunction, the $U-V$ colors show no clear trend.

The long readout time for the detectors meant that individual beat pulses are not well-sampled in the data. But the 15s exposures are short compared with the approximately 2~min beat cycle time, so we have some information on the spectral variations of the beat/spin cycle even while not resolving individual pulses. Figure~\ref{prom_c} shows short timescale variations in the H$\alpha$ emission line around orbital phase 0.75. Rapid emission variations are seen near phase 0.25 as well, but a gap in the data prevents fully exploring this part of the orbit. 

These flashes of emission are most obvious around quadrature, when the secondary star is seen at its highest red or blue shifted velocities. This is also the phase when the satellite emission lines we identify with prominences are seen at their lowest radial velocities. It is not clear if the fast time-scale emission flashes are invisible at phases away from secondary quadrature, or if the prominence emission prevents their detection.

\begin{figure}[]
\includegraphics[width=0.5\textwidth]{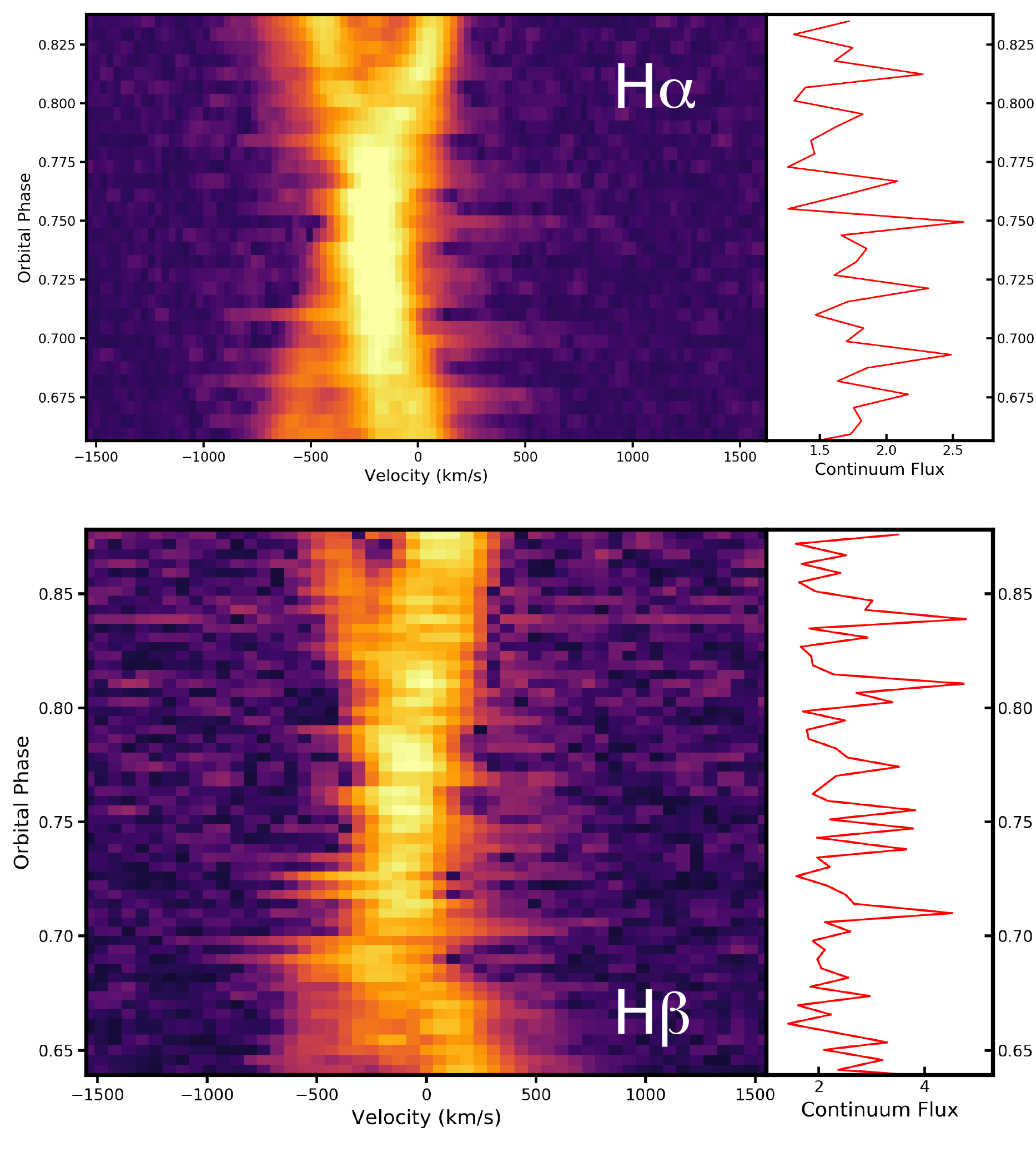}
\caption{The stacked H$\alpha$ and H$\beta$ spectra zoomed in on orbital phase $\sim$0.75. A median average of the spectra was subtracted from each exposure to allow faint features to stand out. Flashes of high-velocity emission are seen extending to 700~km$\ $s$^{-1}$ from the secondary. The emission alternates between red and blue shifted velocities on a time scale of the beat/spin period. The plots to the right of the images show the continuum light variations matching the phases of the emission spectra. In general, the redshifted flashes correspond to bright pulses in the continuum light while blue shifted flashes tend to be associated with the continnum being faint. }
\label{spikes}
\end{figure}

We show stacks of the H$\alpha$ and H$\beta$ lines (Figure~\ref{spikes}) after zooming in on phase 0.75 and subtracting a median line profile to make the varying emission more clearly visible. The flashes of Balmer emission are seen to extend to $\pm 700$~\kms . The emission flashes appear to alternate in the red and blue shifted directions. Comparing with the continuum flux variations, the red shifted flashes appear to occur when the continuum is bright, while the blue shifted flashes prefer moments when the continuum is at its faintest.

To quantify the timing of the emission line flashes, we applied the ``double Gaussian'' \citep{shafter83,schneider80} technique to the H$\alpha$ and H$\beta$ lines. The method consists of convolving an emission line with two Gaussian functions separated in wavelength to avoid the bright core of the line. The wavelength where both Gaussian functions contain equal flux is an estimate of the velocity centroid of the highest velocity emission. Normally, the double Gaussian method is applied to disk systems to approximate the motion of the white dwarf star where the highest velocity emission corresponds to the disk gas closest to the primary. Here, we aim to avoid the brightest emission in the lines and detect centroid shift caused by the emission flashes. 

For H$\alpha$ and H$\beta$, we used Gaussian widths (sigma) of 130~\kms\ and a full separation of 1400~\kms . The separation velocity was chosen to maximize the detection of the line flashes which are seen to extend out to 700\kms\ from the line center. The resulting radial velocity curve (Figure~\ref{shafter}) shows the overall motion of the secondary and satellite lines, plus rapid velocity fluctuations around quadrature with amplitudes of 50 to 100~\kms . 

We repeated the double Gaussian analysis for the H$\beta$ line, removed the sinusoidal variations and combined the velocity variations observed in the two Balmer lines. Because the red and blue side of the spectrograph were running independently, the cadence of the velocity measurements was significantly improved when the two curves were intertwined. The average time between spectra was 0.51 minutes after the red and blue sides were combined. A power spectrum of the velocity variations showed a strong peak in the velocity variations at 0.980~min, which is half the beat/spin period. Caution must be used in this analysis as the periodicity we see in the Balmer emission flashes is close to twice the sampling rate.

\section{Discussion}

\subsection{Prominences}

From the H$\alpha$ tomogram, we can estimate the size and location of the satellite emission that we associate with prominences by assuming that they are co-rotating with the secondary star. The projected distance of these features from the center of mass (CM) of the binary is simply $r=v/\Omega$, where $v$ is the inclination-corrected velocity amplitude of the emission and $\Omega$ is the angular velocity of the secondary. The orbital inclination is expected to be $\sim75^\circ$ based on our assumed stellar masses and the mass function measured by \citet{marsh16}. Therefore, the correction between the radial velocity space velocity is less than 10\%. For the orbital period from \citet{marsh16}, the inverse angular velocity is 2043~s$\;$rad$^{-1}$. Assuming a mass for the WD of $0.8$~M$_\odot$, then the mass ratio of the system is  $q\approx 0.3$ for a secondary with a mass typical for an M5 dwarf. The radius of the secondary's Roche lobe is $R_{RD}\approx 2.3\times 10^{5}$~km  \citep{egg83}. Since the secondary is probably close to filling its Roche lobe, we use this as the stellar radius and compare it to the size of the prominences. The brightest emission from the prominences is seen around 500~\kms, corresponding to 4.4 Roche radii from the center of mass. The prominences precede and trail the secondary by $\pm$0.15 in phase, meaning that the prominences extend 3 stellar radii from the surface of the secondary. The faint emission behind the secondary at 1000~\kms\ corresponds to 8.7 stellar radii from the CM, or about 5 stellar radii from the backside of the red dwarf.

In isolated magnetized stars, the co-rotation radius is where the gravitational pull of the star balances the centrifugal acceleration of plasma trapped in the field. Prominences found at or beyond the co-rotation radius are called "slingshot" prominences \citep{slingshot}. In binary systems the balance between centrifugal and gravitational forces is found at the Lagrangian points. The L$_4$ and L$_5$ points are actually rather broad regions that make an equilateral triangle with the primary and secondary stars. For the assumed parameters of the AR~Sco binary, the L$_4$ and L$_5$ points are 370~\kms\ from the CM. The prominences extend more than 30\%\ farther than L$_4$ and L$_5$ points. The trailing prominence differs in phase from the L$_5$ point by 20$^\circ$.

The L$_2$ point is expected to be at 4.5 Roche radii from the CM, or a velocity of 500~km/s, while the faint prominence behind the secondary is seen at twice that distance. We conclude that the prominence emission extends well beyond the outer Lagrangian points.

The stability of slingshot prominences has been modelled by \citet{ferreira00}. Such prominences can be stable out to $\sim 5$ stellar radii if the dipole surface magnetic field is larger than $\sim 100$~G, and out to 10 stellar radii for a 500~G field. A quadrupole field would require $\sim 500$~G for prominence stability at 5 stellar radii. 

The WD field at the secondary is expected to have an average field strength of order $\sim$200 Gauss \citep{takata18}, so the magnetic fields from the two stars should be comparable just at or above the red dwarf surface. Thus, the energy produced in the AR~Sco system is likely the result of the interaction between the two star's magnetic fields.

\subsection{Emission and Continuum Light Curves}

\begin{figure}[]
\includegraphics[width=0.5\textwidth]{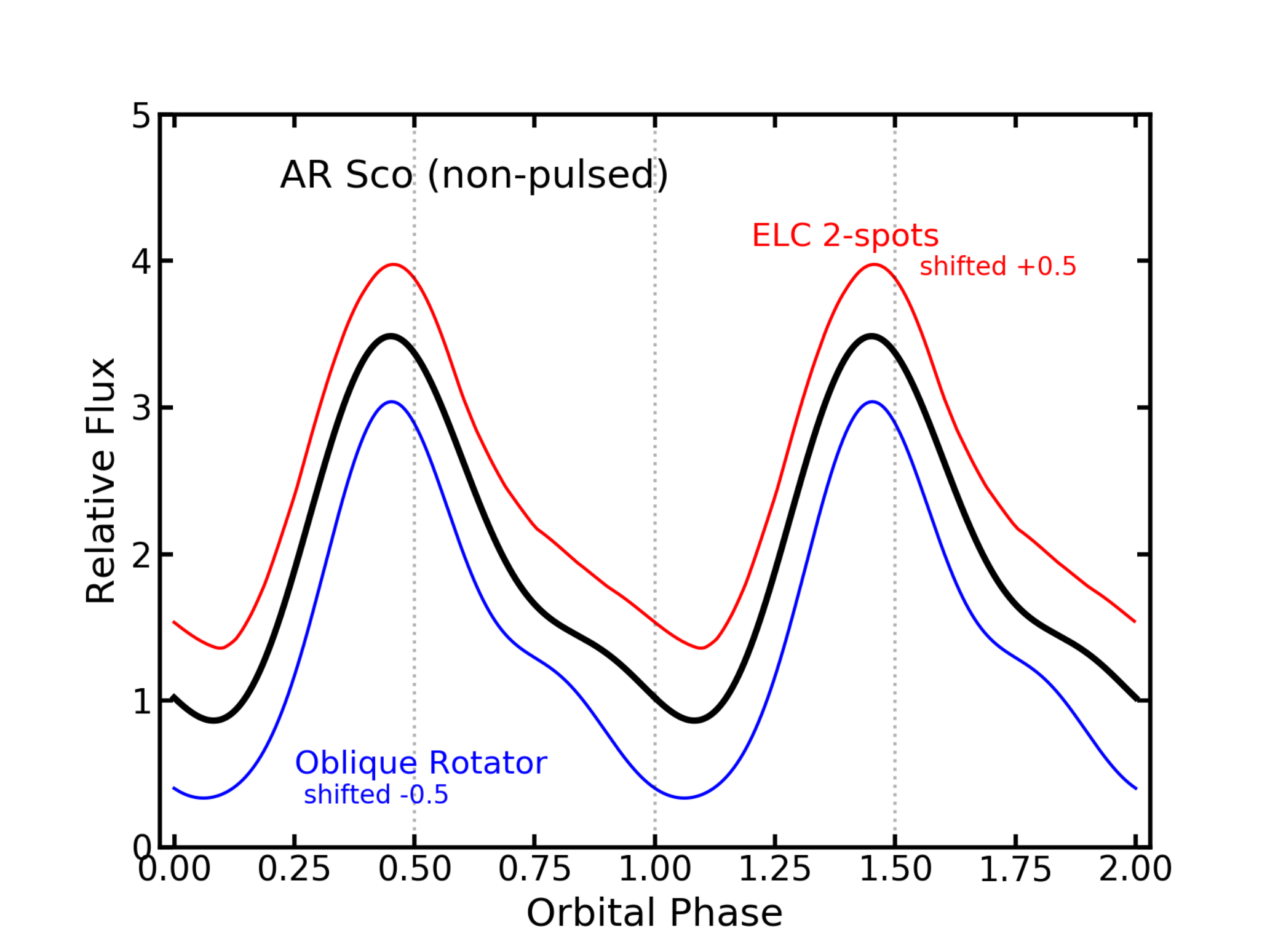}
\caption{The non-pulsed optical light curve (heavy black line) from \citet{stiller18} is compared with the two-spot ELC model (red line) and the oblique WD rotator model (blue line). The orbital phase is based on the radial velocity measurements of the secondary \citep{marsh16} where phase 0.5 corresponds to binary superior conjunction. Both models can match the orbital phase shifts observed in the minimum and maximum light from the system.}
\label{model_lightcurves}
\end{figure}

 We use the light-curve-modelling program ELC \citep{ELC} to create a toy model of the light variations on an M5 star filling its Roche lobe. The strong beat pulses are likely not coming directly from the secondary star, so we use the light curve created by \citet{stiller18} that excludes pulsed emission by selecting only the lower envelope of the high-cadence flux measurements. The \citet{stiller18} light curve has the flux minimum shifted from inferior conjunction by nearly 0.1 in phase and the maximum occurring before superior conjunction by 0.05 in phase. This is similar to our continuum- and emission-line-flux curves in Figure~\ref{flux}. Thus, the rise from minimum to maximum takes only 35\% of an orbit. ELC can model this behavior by adding two warm spots to the surface of the secondary (see Figure~\ref{model_lightcurves}). This may not be a unique solution, but the addition of warm spots does an adequate job of matching the light variations by shifting the times of minimum and maximum. 

Spot~1 is on the trailing inward-facing quadrant of the secondary and is significantly hotter than the majority of the M5 star.\footnote{The temperature and area of the warm spots are degenerate in the model, so we refrain from quantifying the temperature.} This is the quadrant in which we expect energy deposition from the magnetic field interaction. By its location on the trailing side adjacent to the L$_1$ point, Spot~1 becomes fully exposed to our view just before superior conjunction. The location of this warm spot in the ELC model results in the peak of the light curve occurring at an orbital phase of 0.45.

The shift in phase of the continuum-flux minimum is not well explained by the ELC model. For the adopted inclination angle of 75$^\circ$ (see Section~4.1), the warm spot on the trailing inward face is almost completely self-eclipsed by the secondary. The observed minimum at phase 0.1 can be achieved by adding a warm spot on the outward-facing, leading quadrant with a modest temperature increase  above the ambient surface temperature. A physical motivation for a warm spot on the backside of the companion is given by \citet{romani16} in modeling the light from "black widow pulsars". They argue that the magnetic field of the secondary star may channel material and energy from the interaction region to the far-side of the cool star in these pulsar systems. This would locally heat the photosphere at the magnetic field line foot points. Our direct evidence of a large-scale magnetic field on the secondary star suggests this "ducting" is possible in the AR~Sco system as well.

For an alternative explanation for the shape of the light curves in AR~Sco we can consider a small tilt to the spin axis of the WD relative to orbital plane, also called an obliquity. Such a scenario was proposed by \citet{katz17}. For simplicity, we can further assume that the star is an orthogonal rotator, with its magnetic poles near the equator of the WD. With these assumptions, any interaction between the magnetic field of the WD and the secondary star will be maximized at the two orbital phases where the magnetic axis crosses the orbital plane. At very high obliquities the interaction producing the beat pulses would be isolated to these two phases, but this is not what is observed for AR~Sco. Instead, a mild obliquity would modulate the interaction sinusoidally at twice the orbital frequency. A bright spot near the L$_1$ point generated by the magnetic interaction would be seen to have two maxima and minima per orbit. Combine this modulation with the typical irradiated spot being viewed around an orbit and there are sufficient free parameters to match the observed light curve variations.

For a simple example we imagined a hot spot at the L$_1$ point whose viewing angle varies around the orbit. Its light curve is approximately a sinusoid with a maximum at the secondary star's superior conjunction. We then multiply by a sinusoid with twice the orbital frequency to represent the inferred interaction modulation due to an obliquity. The modulation is offset in phase so that the maxima occur when the WD equator crosses the orbital plane (the "equinox" if you will). Using a modulation amplitude of 20\%\ shifted by 30$^\circ$ in orbital phase provides a fairly good match to our observed continuum and emission line flux curves as shown in Figure~\ref{model_lightcurves}. One sees a rapid rise to maximum taking only 40\%\ of the orbit followed by a slower decay. 


\begin{figure}[]
\includegraphics[width=0.5\textwidth]{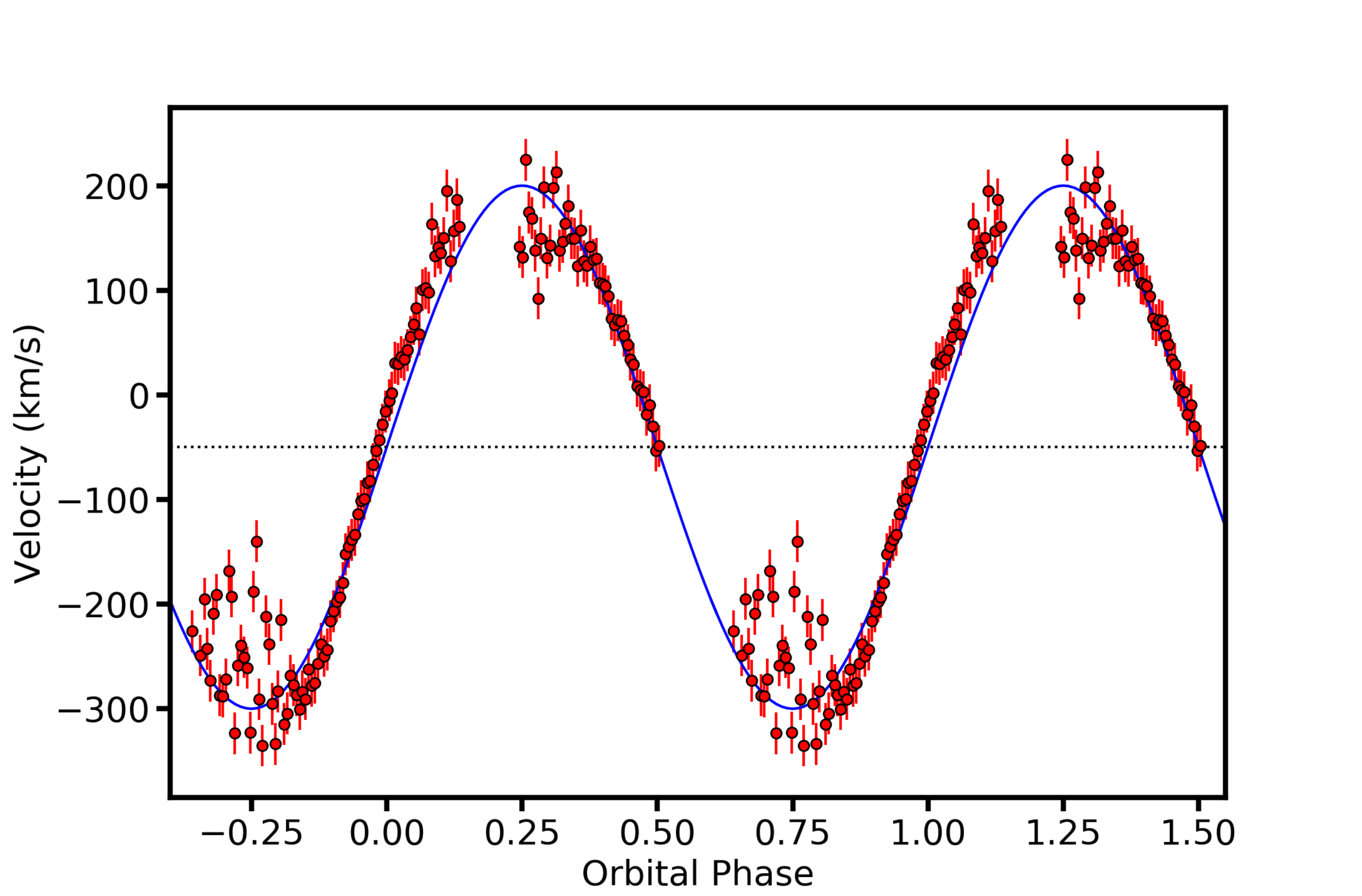}
\caption{The H$\alpha$ emission line velocity centroid as measured from the wings of the line using a double Gaussian technique. The large scatter seen at the extreme velocities is caused by the emission line flashes occurring at the beat/spin period. The satellite emission dominates the other phases and likely distorts the radial velocity curve. The solid line is a sinusoid with an amplitude of 250~\kms\ and phased to the Marsh orbital ephemeris. }
\label{shafter}
\end{figure}

\subsection{Location of Emission and Absorption Lines}

As noted by \citet{marsh16}, there is significant emission line flux coming from the hemisphere of the secondary star facing the WD. Such emission is not uncommon in CVs due to the irradiation of the secondary from energy generated by accretion onto the WD. In AR~Sco however, WD accretion is not the major source of ionizing radiation, and emission lines coming from the secondary may provide information on the source of the hard radiation.  The H$\alpha$ tomogram made from our spectra (Figure~\ref{tomogram}) confirms that strong Balmer emission originates near the L$_1$ point at an orbital velocity $\sim 220$~\kms, and while it suggests that the H$\alpha$ emission may be more prominent on the leading side of the star than the trailing side, the difference is not definitive. 

\begin{figure}[]
\includegraphics[width=0.30\textwidth]{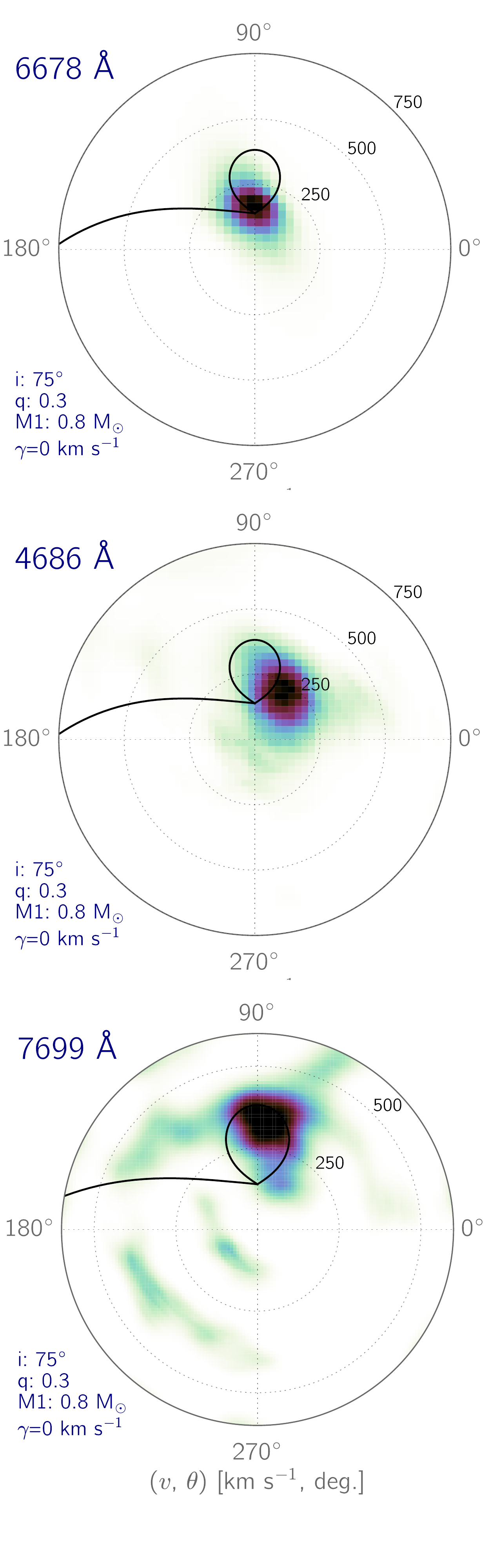}
\centering





\caption{Standard tomograms of three spectral features with a range of excitations. {\bf top:} The distribution of He~I (6678 \AA) emission showing it symmetrically placed near the L$_1$ point of the secondary star. {\bf center:} A half-orbit tomogram of He~II (4686 \AA), showing the distribution to be concentrated near the trailing face of the secondary star. {\bf bottom:} The strength of K~I (7699 \AA) absorption indicating that its presence is mainly on the far side of the secondary away from the WD. The K~I absorption appears to avoid the leading side of the secondary suggesting a warming there.  }
\label{tomos}
\end{figure}

A warm spot in the trailing face of the red dwarf suggests an asymmetric irradiation of the secondary from the source of the synchrotron emission. We have computed tomograms of individual spectral features using the code from \citet{kotze} to further test for asymmetric irradiation, and these are shown in Figure~\ref{tomos}.

The HeI 6678\AA\ emission line is brightest on the inward-facing side of the secondary and is fairly symmetric about the L$_1$ point. Its distribution on the secondary is similar to the brightest emission in the tomogram of H$\alpha$ as found by \citet{marsh16}.

The HeII 4686\AA\ line is weak but present in the individual spectra for part of the orbit.\footnote{Because Doppler tomography assumes that line-emitting regions are uniformly visible across the orbit, the HeII tomogram uses spectra from the half of the orbit during which HeII is strongest \citep[e.g.,][]{potter04}.} Its tomogram indicates that HeII emission is concentrated on the trailing face of the secondary and this maybe where much of the magnetic interaction is depositing its energy.

The neutral potassium absorption line at 7699\AA\ is a good tracer of cool gas because of its low ionization energy. The KI tomogram shows little absorption on the inward face of the secondary that is likely irradiated by the interaction. The distribution of the KI absorption strength confirms that the backside is cooler than the inner face of the secondary.

\subsection{Emission Line Flashes}

\begin{figure}[]
\includegraphics[width=0.5\textwidth]{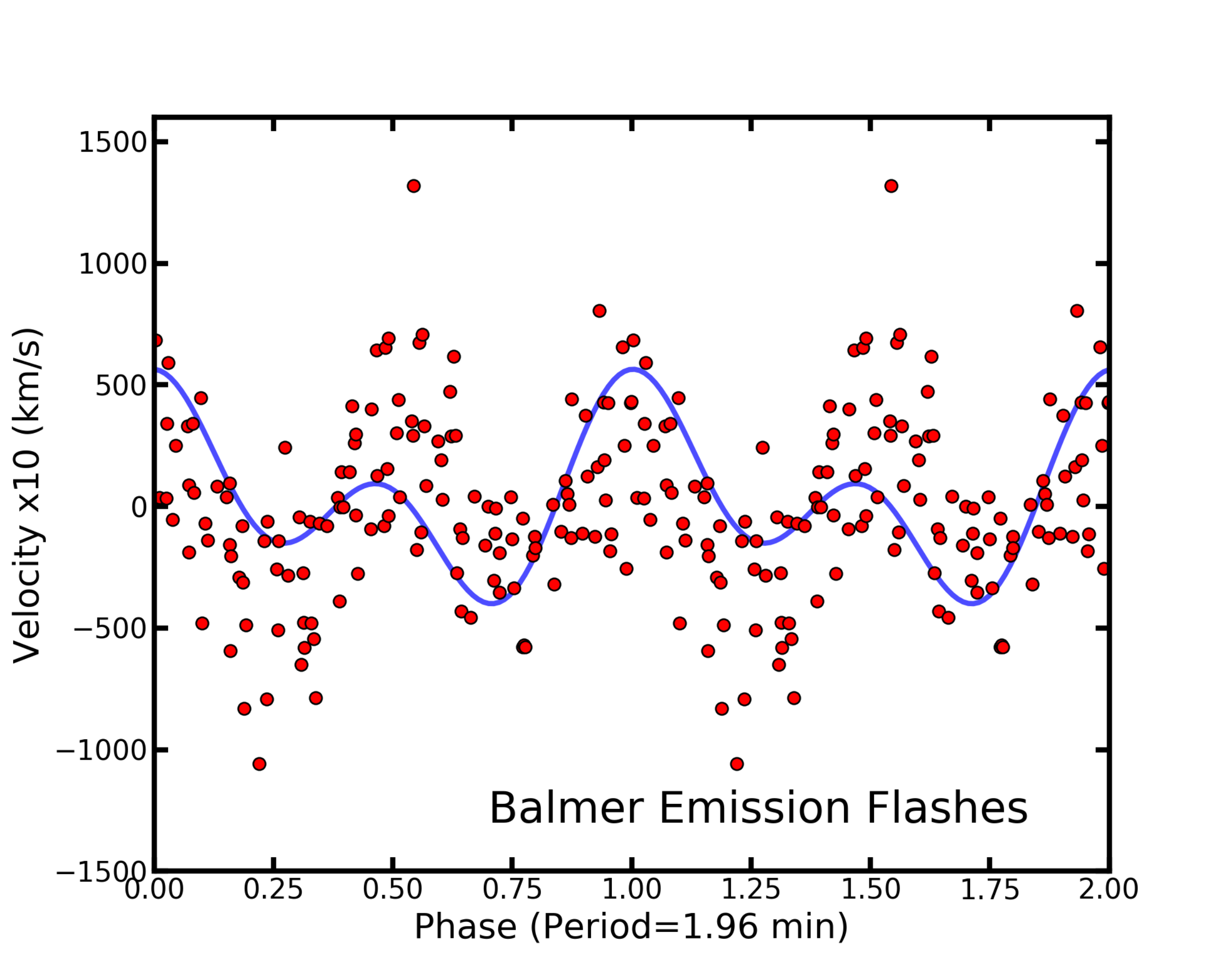}
\caption{The Balmer emission velocity centroid from the double-Gaussian analysis of spectra obtained around orbital phases 0.25 and 0.75. H$\alpha$ and H$\beta$ have been combined to improve the cadence. The circles show the velocity phased on a period of 1.96~min, close to the beat and spin periods. The line shows a photometric model of the beat pulses from \citet{stiller18}. The red shifted emission line flashes tend to occur when the beat pulse is bright while the blue shifted flashes correlate with the beat pulse being faint. }
\label{high_vel}
\end{figure}

The Balmer-emission flashes have not been detected in previous observations of AR~Sco. They are observed around quadrature of the secondary, but that may simply be due to the prominence emission making them difficult to see at other phases. After removing the orbital variations in Figure~\ref{shafter}, we phased the velocity centroid measurements to the beat period and plot the result in Figure~\ref{high_vel}. The red-shifted flashes are correlated with brightening in the continuum, while the blue shifted emission line flashes occur when the continuum is at its faintest level over a beat cycle. Because these are centroid shifts, the double Gaussian amplitudes are about a factor of ten smaller than the maximum velocities seen of the emission line flashes.

The correlation between the velocity of the flash emission and the continuum light variations is surprisingly good considering the poor cadence of these observations and the limited orbital phase coverage of the flashes. More data, over several orbits, are needed to confirm the correlation seen in Figure~\ref{high_vel}.

Our data shows hydrogen emission flashes on time scales of the WD beat/spin period and velocities as high as $\sim 700$~\kms\ relative to the secondary star. The mechanism that accelerates hydrogen to these velocities and still allows for Balmer emission is not clear. It is unlikely that protons would be accelerated in magnetic fields, then recombine with electrons to generate the observed Balmer emission. Although, charge exchange with neutral hydrogen may occur under the correct conditions. 



\subsection{Implications}

The presence of prominences in the AR~Sco system implies that the magnetic field of the secondary is comparable to that of the WD near the L$_1$ point of the system. The rapidly spinning WD field may generate a time-dependent interaction between the two magnetic systems resulting in magnetic field reconnection on a regular 1-minute time-scale. These magnetic reconnection events produce solar flare-like bursts of energy that generate the observed synchrotron radiation. The secondary star would absorb some of the energy from these flares and be heated around its inner face. 

\begin{figure}[]
\includegraphics[width=0.5\textwidth]{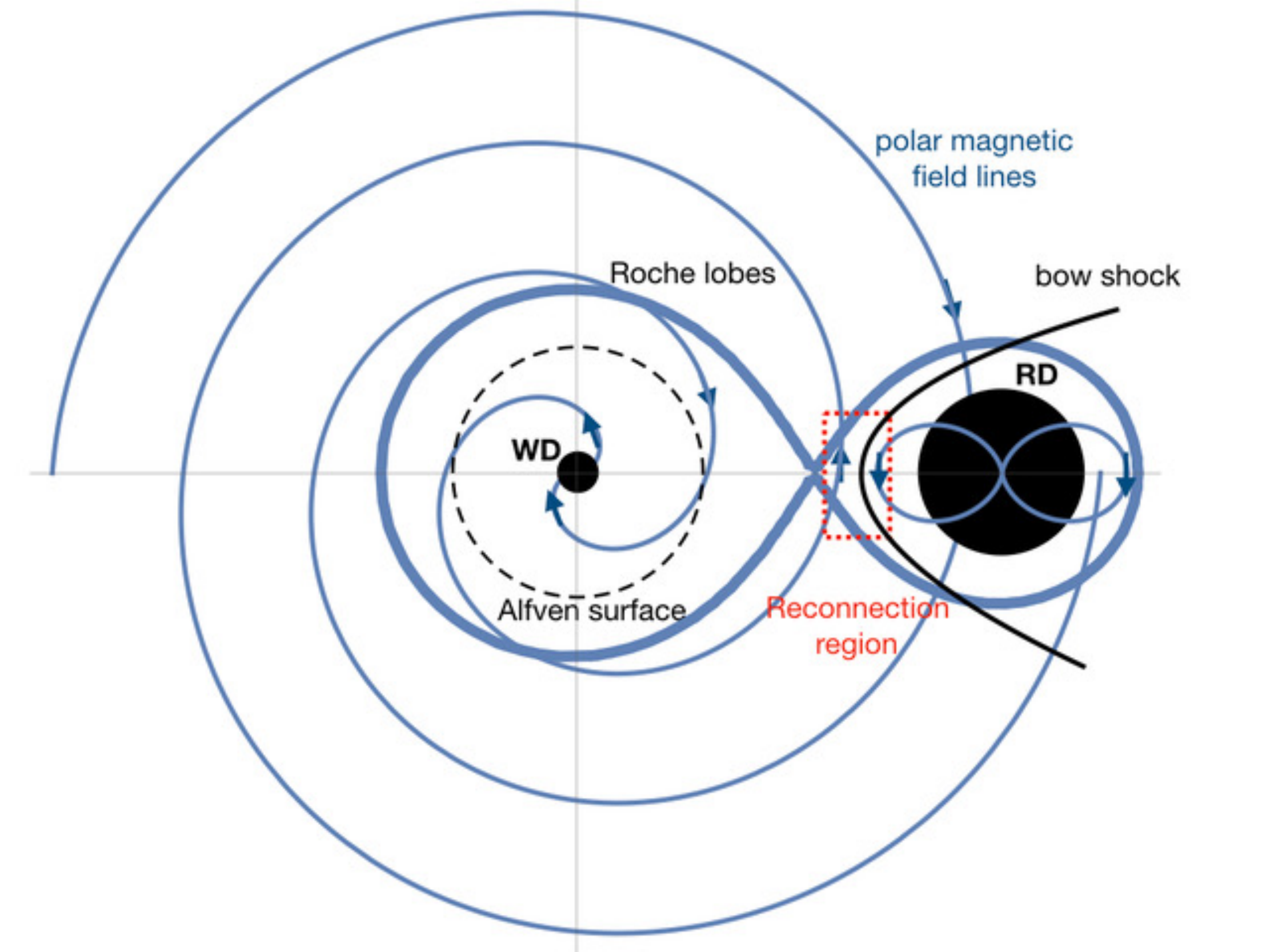}
\caption{A sketch of the AR~Sco system suggesting how the interacting fields of the two stars might generate emission and jets in a magnetic reconnection region. The view is from above the orbital plane and the distances in the diagram are not to scale. The interaction likely occurs close to the red dwarf surface resulting in an irradiated region on the inward face of the secondary.  }
\label{model}
\end{figure}

Here we estimate the energy that could be generated in this scenario. A more detailed model will be presented in \citet{lyutikov19}. We assume the WD is close to being an orthogonal rotator. The  direction of the toroidal \Bf\ in the WD magnetosphere flips twice every rotation of the WD, leading to periodic reconnection events, see Figure.~\ref{model}.

The value of the \Bf\ of the WD at the location of the red dwarf depends on the \Alfven velocity in the WD magnetosphere. Within the \Alfven radius the field falls off rapidly ($B \propto r^{-3}$), while outside the strength falls off more gently ($B \propto r^{-1}$). The Alfven velocity in the magnetosphere of the white dwarf is not known.
For an order-of-magnitude estimate we assume that \Alfven radius is about the size of the Roche lobe, so that the dipolar field of the WD at the L$_1$ point is given by 
\begin{equation}
B_{L_1} = B_{WD} \left( \frac{R_{Roche}}{R_{WD}} \right) ^{-3}
\end{equation}
where $B_{WD} $ is the surface field, and $R_{Roche} \approx 6 \times 10^{10}\; {\rm cm}$ is the radius of the Roche lobe of the WD. For a surface \Bf\ of $ B_{WD}= 500$~MG and a typical WD radius, then the field in the interaction region is about 300~G, similar to that estimated by \citet{takata18}.  

Thus, we expect reconnection events between the two \Bfs\ to occur when the fields are counter-aligned in the interaction region. The power from reconnection is the rate at which the magnetic energy within the reconnection volume is lost. We expect the interaction to occur near the inner face of the secondary, so a good estimate of the size of the interaction region is of order the radius of the secondary star, $R_{RD}\approx 2\times 10^{10}$~cm. The rate at which the reconnection can take place is defined by the \Alfven\ velocity given by
\begin{equation}
 {v_{A}} = \sqrt{\frac{B^2}{ 4\pi \rho}}
\end{equation}
where $\rho$ is the plasma mass density in the reconnection region. The electron density in the Solar chromosphere \citep{solarflare} is typically estimated to be 10$^{11}$ to 10$^{12}$~cm$^{-3}$.  The electron density measured in stellar flares \citep[e.g. AB~Dor;][]{stellarflare} is also in that range. These electron densities roughly correspond to a mass density of $\rho\sim 10^{-12}$~g$\;$cm$^{-3}$. For a magnetic field of 300~G, the \Alfven\ velocity is then $v_A \approx 10^{8}$~\cms.

Given the above estimates, we find the reconnection power is
\begin{equation}
L_{rec} \sim \frac{B^2}{8\pi} \pi R_{RD}^2 v_A 
\sim 6  \times 10^{32}  {\rm erg s}^{-1}
\label{Lrec}
\end{equation}
 which is comparable to the excess luminosity measured from AR~Sco \citep{marsh16} and only slightly less than the total power available from the WD spin-down \citep{stiller18}.
 
 The rapid reconnection events should generate ``reconnection jets'' \citep{recon-jet} with velocities similar to the \Alfven\ speed. Our rough estimate of an \Alfven\ velocity $v_A \approx 10^3$~\kms\ is comparable to the velocities we observed in the Balmer emission line flashes and these flashes may well be explained by reconnection jets. However, the jets will accelerate charged particles and it is not certain how neutral hydrogen or recombination emission can result directly from the jet.  An examination of the ion-neutral coupling coefficients from \citet{draine86} or \citet{draine83} makes it plausible that some of the neutral gas would also couple to the ionized gas; but we leave a detailed examination to \citet{lyutikov19}.
 
 In summary, magnetic reconnection events occurring near the L$_1$ point can account for the power observed from the AR~Sco system using reasonable parameters for the magnetic field, density, and size of the interaction region. Such reconnection events, like stellar flares, generate synchrotron emission that can have a short cooling time. The location of the interaction naturally explains the ionization seen on the secondary. Further, particles accelerated during magnetic reconnection may account for the high velocity emission flashes we detect near secondary quadrature.

\section{Conclusion}

Moderate resolution spectroscopy of the white dwarf pulsar AR~Sco shows the presence of magnetic loops or prominences originating from the red secondary star. The prominences extend 3 to 4 Roche radii from the trailing and leading sides of the secondary. There also appears weak emission from a loop extending 5 Roche radii on the back side of the red star. The spectra also show:

\begin{itemize}

\item The prominence emission is strongest in H$\alpha$ and due to a steep decrement the becomes less prominent in higher order Balmer lines.

\item Beat/spin pulses dominate the continuum variations, but line emission strengths show little change during the beat/spin pulses.

\item The strength of the emission lines (excluding prominence emission) reaches a minimum between orbital phase 0.05 and 0.10, as does the continuum.

\item The average $U-V$ color is reddest just after orbital phase zero while the average $B-V$ color does not strongly vary over the orbit.

\item Based on tomography, the HeII emission comes from close to the trailing, inner face of the secondary, suggesting significant energy deposition at that location. In contrast, the neutral potassium line from the red dwarf is found only on the outward face of the secondary 

\item The peak of the beat pulse shows the bluest colors and the peak color is consistent with synchrotron emission. As the flux fades from the pulse peak the continuum reddens with a slope consistent with synchrotron cooling.

\item Flashes of Balmer emission are seen extending  $\pm 700$~\kms\ during quadrature of the secondary. 

\item The red-shifted emission flashes appear to be synchronized with the bright continuum pluses. The blue-shifted flashes is timed with the faint phases of the beat emission. Sampling of the timing is limited and further data is needed to confirm this relationship.

\item Magnetic reconnection can account for the energy production in AR~Sco.  The location of the reconnection can explain the heating and ionization on the inner face of the secondary. Reconnection jets are a natural possibility for the source of the high velocity flashes seen in our spectra.

\vspace{0.5cm}

\end{itemize}

The existence of long-lived slingshot prominences emanating from the secondary in AR~Sco suggests a surface magnetic field between 100 and 500~G. As the magnetic field of the WD is expected to be about 300~G near surface of the secondary, we propose that the luminosity from AR~Sco is generated by the magnetic reconnection between the two magnetic fields. The interaction occurs just above the surface of secondary and on its trailing side. The resulting synchrotron radiation heats that region of the secondary and generates the irradiated emission features seen in the spectra.

\acknowledgments We thank Josh Walawender, Cindy Wilburn, and the Keck Observatory for help in obtaining these observations. DSB acknowledges support via NSF grants NSF-ACI-1533850, NSF-DMS-1622457, NSF-ACI-1713765, and NSF-DMS-1821242. Support from a grant by Notre Dame International is also acknowledged. We thank L. A. Phillips for insightful comments. MRK is funded by a Newton International Fellowship from the Royal Society.

\facility{Keck Observatory} 


\end{document}